\documentclass[12pt]{spieman} 
\usepackage{amsmath,amsfonts,amssymb}
\usepackage{graphicx}
\usepackage{setspace}
\usepackage{tocloft}
\usepackage{soul}
\usepackage{booktabs}
\usepackage{wrapfig}
\usepackage{multirow}
\usepackage{changepage}
\usepackage{wrapfig}

\title{Great Observatories Maturation: a Review of NASA Astrophysics Development Through Suborbital Rocket and Balloon Programs}

\author[a*]{Drew M. Miles}
\affil[a]{The California Institute of Technology, 1200 E. California Blvd., Pasadena, CA 91125}

\cftpagenumbersoff{figure}
\cftpagenumbersoff{table} 
\begin{document} 
\maketitle

\begin{abstract}
The NASA Great Observatories Maturation Program is a development plan to efficiently and effectively develop large, strategic astrophysics missions. Suborbital rocket and balloon programs have long been a key development tool for enabling large missions in NASA astrophysics. We review the significance of these suborbital missions in the preceding decades to demonstrate their contributions to the Great Observatories Maturation Program for the Habitable Worlds Observatory and beyond.  
We show that suborbital instruments have obtained new science observations of astrophysical sources across the electromagnetic spectrum, matured high-priority component technologies, and served as a training ground for principal investigators of Explorer-class astrophysics satellites. A brief discussion of emerging CubeSat and SmallSat missions and their place in the NASA astrophysics portfolio is also provided. 

\end{abstract}


{\noindent \footnotesize\textbf{*}Drew M. Miles,  \linkable{dmiles.astro@gmail.com} }

\begin{spacing}{1}  

\section{Introduction}
\label{sec:intro}  

A primary recommendation from the 2020 decadal survey in astronomy, \textit{Pathways to Discovery in Astronomy and Astrophysics for the 2020s} (Astro2020), is a large-scale program that provides ``significant early investments in the co-maturation of mission concepts and technologies for NASA large strategic missions'' \cite{Astro2020}. Designated as the highest priority for enabling space programs in Astro2020, such a program is designed, in part, to increase the frequency of large strategic missions in NASA astrophysics\cite{Astro2020}; just one such mission has launched this millennium: the James Webb Space Telescope (JWST), borne from the highest-priority recommendation in the 2001 counterpart to Astro2020 \cite{Astro2001}. 
Though the survey does not provide a detailed handbook for implementation, several specific recommendations are included to support the suggested program. The development of mission-specific technical capabilities, subsystem-level demonstrations, and multi-functional teams of scientists and technologists are each identified as integral components to an effective maturation program \cite{Astro2020}.
Further, the decadal recommendation identifies smaller projects (specifically Explorer-class NASA missions) as a key input into the program, highlighting the capability of such missions to collect data at wavelengths not accessible with other observatories and to mature instrument architectures for larger, strategic missions (also known as Flagship missions, refer to e.g. Figure 7.3 of Astro2020).\cite{Astro2020}

The emphasis on this ``Great Observatories Mission and Technology Maturation Program'' represents an evolving approach to mission maturation compared to prior decadal surveys in astronomy; the 2001 survey prioritized specific missions for all recommended major and moderate space-based initiatives \cite{Astro2001}, and the 2010 survey recommended medium-scale, targeted technology development activities to supplement the high-priority large-scale missions \cite{Astro2010}. 
Rather than develop specific technologies for future mission concepts, as with the recommendations from the 2010 survey, Astro2020's recommended program is aimed at ``designating appropriate scope at an early stage and making significant investments in maturing missions prior to the ultimate recommendation and implementation''\cite{Astro2020}. 
Such an approach can allow for a perpetual program that feeds development for large missions with advanced capabilities to operate across a range of wavelength bands, enabling simultaneously operating, complementary ``Great Observatories''. 
Indeed, a primary objective of the program is to realize a panchromatic suite of observatories similar to the ones launched near the end of the last century: the Hubble Space Telescope (ultraviolet/optical), the Compton Gamma Ray Observatory\cite{Kniffen89}, the Chandra X-ray Observatory\cite{Weisskopf00}, and the Spitzer Space Telescope (infrared)\cite{werner04}. 

   \begin{wrapfigure}{r}{0.4\textwidth} 
   \vspace{-0.9cm}
   \begin{center}
   \begin{tabular}{c} 
   \includegraphics[width = 0.38\textwidth, trim={12cm 0cm 12cm 0cm}, clip]{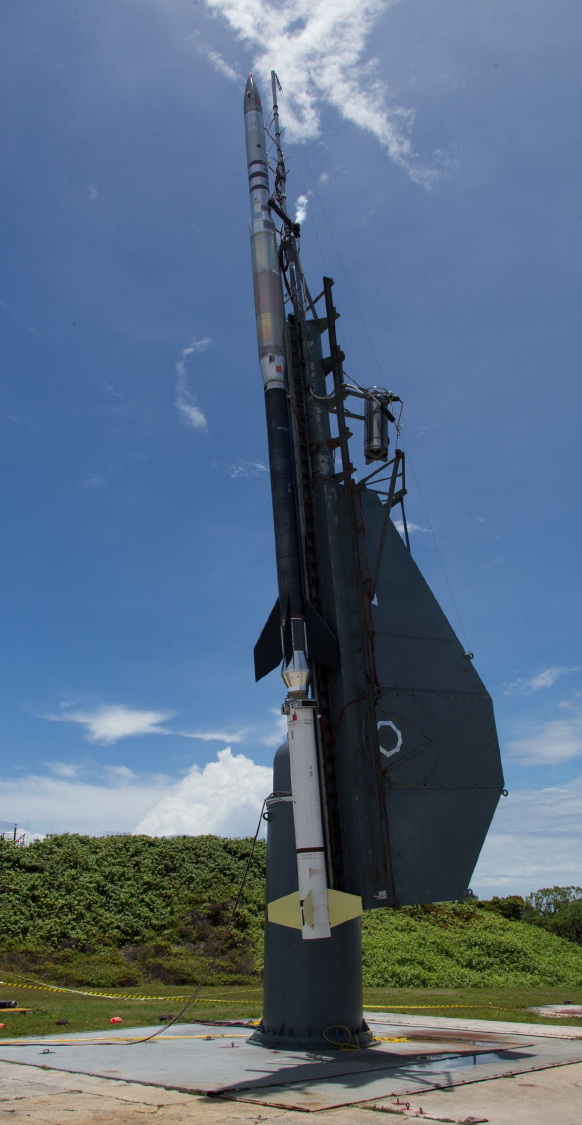}
   \end{tabular}
   \end{center}
   \caption
   { \label{fig:rocket} 
  A suborbital rocket payload mounted to rocket motors on a launch rail. From the bottom up, two rocket motors are affixed to the science payload, visible as the bulbous section $\approx$3/4 of the way to the top, flight support systems (e.g. telemetry and recovery systems), and a nose cone. Image credit: NASA
  }
  \vspace{-0.3cm}
   \end{wrapfigure}

In response to the Astro2020 recommendation, NASA has implemented the Great Observatories Maturation Program (GOMaP) \cite{Feinberg24}. The first large strategic mission that will drive development within the GOMaP is the Habitable Worlds Observatory (HWO), an infrared/optical/ultraviolet telescope with a suite of instruments\cite{Feinberg24}. As part of the GOMaP, two committees were formed to begin initial concept development for the HWO before a traditional NASA project office was created for the mission. These committees, one consisting primarily of scientists and one of engineers \cite{Feinberg24}, organized a broad community effort to develop exploratory analytic cases for the HWO and establish necessary trade studies early on in the concept maturation phase. Though the HWO is the first mission to be developed in the newly establish GOMaP, future concepts are expected to enter the pipeline following the HWO to motivate maturation for subsequent large NASA missions at X-ray and far-infrared wavelengths, for example.

The large-mission maturation program outlined in Astro2020 emphasizes several components that exist presently within NASA's suborbital rocket and balloon programs. In particular, suborbital projects have been used extensively within NASA astrophysics to develop key technologies and instrument subsystems for space missions, enable cutting-edge science investigations, and train multi-functional teams of early-career scientists and engineers. This paper presents a review and discussion of NASA's suborbital rocket and balloon programs within the context of maturing science, technologies, and mission concepts for orbital-class NASA missions such as Explorers and Flagships. Section \ref{sec:suborbitals} introduces NASA's suborbital programs and contextualizes their place in the NASA mission portfolio, Section \ref{sec:sub_dev} details various developments enabled by suborbital missions and their significance to broader NASA astrophysics and the GOMaP, and Section \ref{sec:smallsats} briefly discusses other small mission types. A summary of this review is provided in Section \ref{sec:summary}. 

\section{NASA's Suborbital Programs}
\label{sec:suborbitals}

NASA's suborbital programs can be broadly defined to include true suborbital missions -- flight programs that leave the Earth's surface but do not achieve orbit  -- 
and small, orbital missions such as CubeSats. 
This review will focus on ``traditional'' suborbital programs, sounding rockets and stratospheric balloons, that collectively constitute the suborbital NASA astrophysics missions that are typically funded through the NASA Astrophysics Research and Analysis (APRA) program. A brief discussion on the significance of airborne observatories and CubeSats is presented in Section \ref{sec:smallsats}.  
NASA-supported suborbital projects include flights dedicated to technology advancement, engineering demonstrations, outreach initiatives for undergraduate and pre-university students, and science applications. Such missions are used for investigations in Earth science, heliophysics, and astrophysics that are often only achievable with the custom instrumentation designed for that suborbital mission.

\begin{figure*}[ht]
   \begin{center}
   \begin{tabular}{c}
   \includegraphics[width = 0.9\textwidth, trim={0cm 0cm 0cm 0cm}, clip]{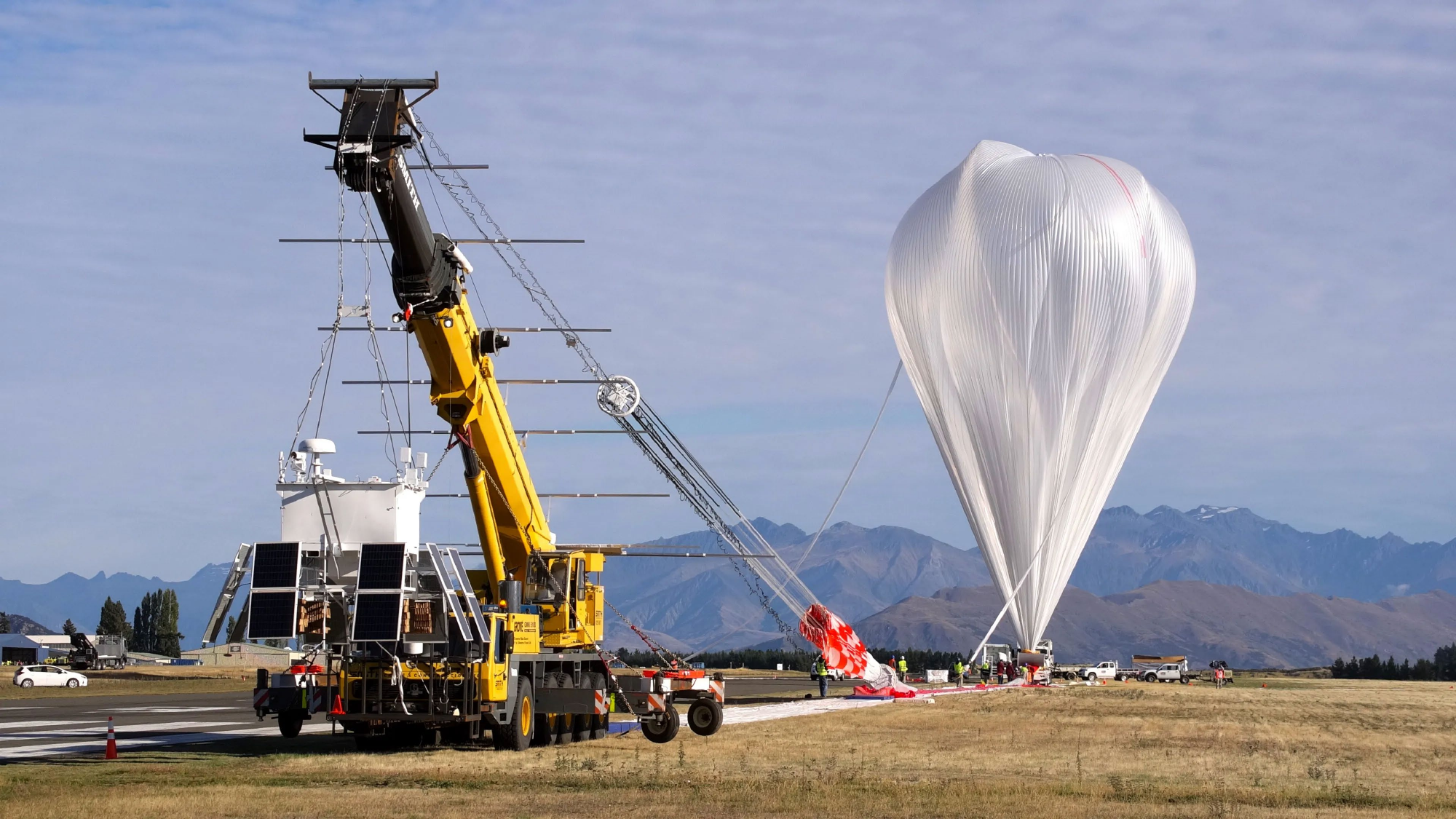}
   \end{tabular}
   \end{center}
   \caption
   { \label{fig:balloon} 
  A super-pressure balloon preparing for launch from Wanaka, New Zealand. The balloon, shown inflated in the background of the image, will support the payload, shown suspended from the launch vehicle in the foreground, at an altitude of $\approx$34~km throughout the flight\cite{moran}.
  }
   \end{figure*}

Table \ref{tab:comparison} shows a summary of key mission characteristics for the primary mission types in the NASA astrophysics portfolio. The approximate mission costs, development time, duration of science mission, launch frequency, and relative risk level of the instrument technologies are compared to demonstrate the wide variance of NASA missions and highlight the unique aspects of each mission type. 
The ``suborbitals'' column has been generalized to include both suborbital rocket and balloon instruments, and each suborbital mission type offers different capabilities in terms of proximity to space and instrument architectures. 
Suborbital rockets, an example of which is shown in Figure \ref{fig:rocket}, can support science payloads up to $\approx$300~kg and offer several minutes of observation time above 250~km in a true space environment. Rocket-borne payloads traditionally offer the fastest development cycle and least expensive method to demonstrating instrument performance beyond Earth's atmosphere. Suborbital balloons (see Figure \ref{fig:balloon}) ascend to and float within the stratosphere, achieving altitudes $\lessapprox$40~km. Balloon payloads can support Explorer-sized instruments (up to several thousand kg) and operate for $\approx$1 day (conventional balloons) or tens of days (long-duration balloons). Comparatively, balloon missions are typically about a factor of two more expensive to realize than suborbital rocket flights and often strive for instrument designs and scales that are more representative of large-mission instrument architectures.

\begin{table}[ht]
\caption{A comparison of the major NASA astrophysics mission types. Time to launch captures the typical development timescale from mission initiation to launch. Technological risk refers to the maturity level of component technologies typically used in science instruments; higher risk implies less mature technologies. We note the absence of Probe-class missions, which are solicited less frequently and are placed between Explorers and Flagships in each of the parameters presented. *Only three astrophysics CubeSats have launched at the time of this review, with a wide variance in development time that is trending longer.  } 
\label{tab:comparison}
\begin{center}       
\begin{tabular}{|c|c|c|c|c|} 
\hline\hline
\rule[-1ex]{0pt}{3.5ex} & \textbf{Suborbitals} & \textbf{Cube/SmallSats} & \textbf{Explorers} & \textbf{Flagships}  \\
\hline\hline\hline
\rule[-1ex]{0pt}{3.5ex}  \textbf{Cost} & $<$\$10M & $<$\$20M & $\approx$\$100M - \$300M & $\approx$\$3B - 10B \\
\hline
\rule[-1ex]{0pt}{3.5ex}  \textbf{Time to launch} & $\approx$2 - 5 yr & $\approx$5 yr* & $\approx$5 - 10 yr & $\approx$20 yr \\
\hline
\rule[-1ex]{0pt}{3.5ex}  \textbf{Mission Duration} & $\lessapprox$days & $\approx$1 yr & $>$2 yr & $>$5 yr \\
\hline
\rule[-1ex]{0pt}{3.5ex}  \textbf{Frequency} & $\approx$3 -- 5 yr$^{-1}$ & $\approx$1 yr$^{-1}$ & every few years & every decade\\
\hline
\rule[-1ex]{0pt}{3.5ex}  \textbf{Technological Risk} & High & Medium/High & Low/Medium & Low \\
\hline
\end{tabular}
\end{center}
\end{table} 

The primary advantages that both types of suborbital projects offer relative to traditional orbital missions (i.e. Explorers and Flagships) are a significantly reduced cost, faster turnaround from conception to launch, the ability to accept higher level of technological risk, and the possibility of iterating on the payload for successive flights (made possible by recovering flown suborbital payloads). The higher risk tolerance for suborbital missions make them the ideal, and often the only, test bed to space-prove cutting-edge technologies and demonstrate their applicability to larger, lower-risk missions. Further, suborbital missions allow for and encourage active leadership roles for students and other early-career scientists. Early-career engagement on the full lifecycle of suborbital missions provides training and experience for those personnel to execute larger missions in the future. An additional benefit of the short development cycles and relatively low cost of suborbital instruments is a higher launch cadence; several astrophysics balloon and rocket payloads are launched each year as opposed to one every few years for Explorers and one every 10$+$ years for Flagship missions. 

An inherent advantage of a higher launch cadence is that suborbital missions are the most responsive way to demonstrate emerging technologies, achieve advancement in growing scientific fields, and address gaps in wavelength coverage from operating astrophysical instruments. Unlike orbital missions, which often freeze their instrument designs years in advance of launch, suborbital missions can integrate emerging technologies much closer to the actual flight and reap the benefits of technology advancements in the form of improved scientific capabilities; many suborbital missions, in fact, are specifically designed around new instrument technologies to realize unique and emerging space-based instrument concepts (e.g. a far-UV spatial heterodyne spectropolarimeter\cite{Harris14}, the image-slicer-based integral field spectrograph on INFUSE\cite{Witt23}, and multi-object spectroscopy with digital micromirror devices on SUMO\cite{Halferty23}).

\section{Suborbital Programs as Development Tools for NASA Astrophysics}
\label{sec:sub_dev}

The significance of NASA's suborbital programs to overall NASA initiatives was studied in a 2010 report by the National Research Council (NRC)\cite{Report10}. That report, which emphasized the need for NASA to increase resource investment in suborbital programs, highlighted the impact that these programs have on NASA research:
\begin{adjustwidth}{50pt}{25pt}
\textit{``Suborbital program elements play vital and necessary strategic roles in NASA’s research, innovation, education, employee development, and spaceflight mission success, thus providing the foundation for achievement of agency goals. The suborbital program elements enable important discovery science, rapid response to unexpected, episodic phenomena, and a range of specialized capabilities that enable a wide variety of cutting edge research.''\cite{Report10}}
\end{adjustwidth}
The report further noted the prominent role that suborbital projects have in developing personnel for both NASA missions and in general: ``Suborbital projects have contributed substantially and disproportionately to the training of
NASA’s technical workforce and more broadly to the nation’s pool of experienced scientists and
engineers.''

More recently, Astro2020 studied and summarized the impact that suborbital programs have on developing technology and skilled personnel. 
The report's conclusions underscore the relevance of suborbital programs, noting that ``NASA’s balloon program plays an important role... for training future generations of technologies and mission leaders,'' and that ``the rocket program provides unique, irreplaceable opportunities for accessing space.'' \cite{Astro2020}
The Panel on Electromagnetic Observations from Space 2 highlighted the importance of NASA maintaining a balanced program of astrophysics missions, including suborbital rockets and balloon experiments, noting that ``suborbital experiments have played a significant role in testing technologies and training researchers, leading to better space missions.'' \cite{Astro2020} 
Astro2020 further emphasized that suborbital rockets are ``crucial for maturing technologies and formally qualifying them for spaceflight, ... and are attractive for developing new PIs and diversifying instrument teams.'' 
The decadal report, like the NRC's 2010 review, consistently highlights the critical role that suborbital programs play in NASA astrophysics development. 

\subsection{Science applications from suborbital missions}
\label{sec:significant}

As far back as the early 1960s, suborbital balloons and rocket platforms have been used to carry instruments for astronomical observations. In the succeeding decades, suborbital flights helped advance astronomical observations and instrumentation as the era of space-based astronomy matured. Early suborbital instruments took advantage of the same benefit that is leveraged by many such missions today: balloons and rockets provided the most accessible means to position telescopes and detectors at altitudes high enough to observe radiation that is otherwise absorbed in the Earth's atmosphere before it reaches the surface. Suborbital rockets and balloons had a substantial impact on the origin and early development of X-ray astronomy, for example; prior to the launch of the first orbital X-ray mission in 1970\cite{Uhuru}, cosmic X-rays were first discovered and studied through a series of suborbital launches throughout the 1960s\cite{Giacconi62, Giacconi64, Grader66, Grader68}. A similar progression drove early gamma-ray astronomy and other wavelengths not accessible from ground-based observatories.  Though the gaps in wavelength coverage for satellite instruments have lessened due to the launch and operation of dozens of orbital astronomy instruments in the past 60$+$ years, suborbital instruments continue to leverage their unique advantages to make advancements in scientific observations across the electromagnetic spectrum. A brief overview of the breadth of scientific discovery enabled by recent suborbital missions follows.

At the highest energies, balloon-borne particle astrophysics instruments such as CREAM\cite{Smith17} and BESS\cite{Abe17} have been used to detect and characterize cosmic rays and measure the low-energy antiproton spectrum. Similarly, Super-TIGER and its predecessor TIGER were cosmic-ray instruments on long-duration balloons that measured Galactic cosmic-ray abundances\cite{Binns14, Murphy16, Walsh23}. 
COSI was a long-duration balloon mission designed for spectroscopic, imaging, and polarimetric studies of gamma-ray emission below $\approx$5~MeV \cite{Kierans16}. Notable COSI results include the detection of 511~keV emission via positron annihilation\cite{Kierans20, Siegert20}, measurements of Galactic diffuse continuum emission \cite{Karwin23}, and polarimetric studies of gamma-ray bursts\cite{Lowell17}. COSI's successes ultimately led to the payload progressing to an Explorer-class mission with the same name, scheduled to launch in 2027 to produce full sky coverage of soft-gamma-ray emission and expand the balloon missions' studies of the positron lifecycle, diffuse gamma-ray emission, and transient events\cite{Tomsick19}. 

In the X-ray regime, hard-X-ray instruments have taken advantage of atmospheric transmission at stratospheric altitudes to fly balloon payloads, including Calibur, HEFT, and HERO. HEFT and HERO emphasized high-performance grazing-incidence telescopes to focus hard X-rays for imaging and spectroscopy\cite{Harrison00,Gaskin14}, and X-Calibur implemented a hard-X-ray polarimeter to measure the polarization of an accreting neutron star \cite{Abarr20}. At the higher altitudes achievable with suborbital rockets, cutting-edge technologies have been used to study the soft-X-ray sky: the XQC and Micro-X rocket programs developed and flew X-ray calorimeters to make new, high-resolution spectroscopic observations of the diffuse X-ray background\cite{Wulf19} and supernova remnants\cite{microx}, respectively.

Several far- and extreme-ultraviolet (FUV and EUV, respectively) rocket-borne spectrographs have been used to study UV radiation and its impact on the evolution of plasmas and systems throughout the Universe. The CHESS rockets used a high-resolution FUV spectrograph to study the interstellar medium along multiple lines of sight toward O- and B-type stars\cite{Hoadley20, Kruczek18}. 
Another FUV spectrograph, SISTINE, was used to characterize the UV radiation environment around exoplanet host stars \cite{Nell24}. Similarly, the DEUCE rocket payloads were sensitive into the EUV and observed stellar spectra to better understand how hot ionizing stars influence the ionization fraction in the intergalactic medium \cite{Erickson21}. 

In 2024, the long-duration balloon mission GUSTO set the new record for an astrophysics stratospheric balloon mission duration with $>$57 days at a float altitude of $\approx$36~km.\cite{gusto24} GUSTO carried a 0.9-m telescope and heterodyne arrays to spectroscopically map the interstellar medium throughout the Milky Way and the Large Magellanic Cloud \cite{Walker22} at terahertz frequencies. GUSTO's long-duration flight built upon a previous version of the payload, the Stratospheric Terahertz Observatory (STO), that flew for 22 days and successfully observed C~\small{II} \normalsize~in the Carina Nebula Complex\cite{Walker10, Seo19}. 
Additional long-wavelength suborbital instruments include the BOOMERanG and MAXIMA balloon payloads, which resolved structure in cosmic microwave background (CMB) anisotropies\cite{Mauskopf00,Hanany00} and provided the first astrophysical measurements to confirm a flat Universe \cite{deBernardis00,Balbi00}; the SPIDER balloon instrument, which constrained CMB B-mode polarization\cite{Nagy17,Ade22}; BLAST, which made several polarizations measurements of molecular clouds\cite{Gandilo16, Shariff19} and detailed the contribution of re-emitted far-infrared light from galaxies to the infrared background\cite{Devlin09}; and CIBER, which observed the infrared background over several suborbital rocket flights. In addition to measuring the diffuse infrared background\cite{Arai15, Matsuura17}, CIBER studied foreground zodiacal light \cite{Korngut22, Takimoto22} and various discrete sources that contribute to the infrared background\cite{Kim17,Cheng21}. Like COSI, the successful CIBER program served as a pathfinder for a subsequent Explorer mission, SPHEREx\cite{Korngut18}.

Though used most commonly to fly instruments designed to detect wavelengths that are not accessible from Earth's surface, suborbital instruments also find success at wavelengths traditionally observed from ground-based observatories. The ANITA missions, for example, used stratospheric altitudes to observe radio emission induced by ultra-high energy neutrinos in Antarctic ice \cite{Gorham19}, and the SuperBIT balloon-borne instrument, a 0.5-m NUV-to-NIR imaging telescope, used an advanced fine-guidance system and the superior astronomical seeing at higher altitudes to achieve near diffraction-limited broadband imaging over a 45-day flight \cite{Gill24}. Further, in addition to their designed applications, suborbital instruments developed for space- and near-space-based observations can be used for relatively fast-response follow ups for targets of opportunity that may not otherwise be achievable. FORTIS, for example, was used to observe UV emission in the tail of the comet ISON\cite{McCandliss16}, and several suborbital instruments have been used to study enhanced solar activity\cite{foxsi,hi-c} and supernovae\cite{Aschenbach87,Bunner88}.

\subsection{Suborbital development for key technologies}

The NASA Astrophysics Division (APD) technology gaps demonstrate the impact that suborbital projects have on technology and instrument maturation for large NASA missions. Solicited and updated every two years as part of the APD's bienniel technology reports (ABTRs), technology gaps are defined as the difference in technology between the current state of the art and the capability needed to enable or enhance future strategic astrophysics missions. 
The described gaps capture key technologies across the electromagnetic spectrum, but the focus here is the technologies that are driven by needs for the HWO as defined in the gap descriptions\cite{ABTR24}. 

Table \ref{tab:tech} summarizes the HWO-driven technology gaps identified in the most recent ABTR, where several of the listed gaps combine additional, more specific gaps in the actual report. The identified gaps lay out the need to develop almost the entire suite of technologies applicable for a mission such as the HWO, including coatings to enhance mirror performance, spectrograph technologies, bandpass and multi-object selection techniques, and high-performance detectors. From inspection, the description of each gap indicates that many technologies have reached their current maturation state via application on suborbital instruments. High-performance UV detectors, including both micro-channel plate (MCP) and silicon-based detectors, for example, have advanced over the past 10+ years via suborbital programs such as FIREBall \cite{Jewell15} and CHESS \cite{Hoadley20}, and the FORTIS rocket program has developed micro-shutter arrays (MSAs) as a multi-object selection technology through multiple flights for over a decade \cite{Fleming13}. 

Further, each of the technologies listed in the current strategic technology gaps are part of active suborbital programs that have flown in the past several years and/or are planned to be demonstrated via multiple currently funded suborbital programs, as indicated in the right column of Table \ref{tab:tech}. In most cases, these suborbital programs are capable of new and/or improved scientific capabilities specifically because they develop and implement strategic technologies and prioritize development for large missions; INFUSE, for example, used an image slicer with a suite of pupil gratings and large-format MCP to fly the first FUV integral field spectrograph (IFS) and obtain new spectroscopic data from shock fronts in supernova remnants \cite{Witt23}. Similarly, the SHIMCO rocket will combine an all-reflective spatial heterodyne spectrometer and electron-multiplying CCD (EMCCD) to map the H$_I$/H$_2$ boundary layer in nearby star-forming regions \cite{Corliss22}. 
A more comprehensive list of key technologies that were developed via suborbital programs and subsequently infused in orbital instruments throughout the history of NASA astrophysics is provided in the NRC's 2010 report \cite{Report10}.

\begin{table}[t]
\caption{Strategic astrophysics technology gaps identified as critical for the HWO, from the 2024 ABTR\cite{ABTR24}. Note that several individual gaps have been combined for this table (e.g. multiple high-performance UV detector gaps combined into a single row), and the list of funded suborbital demonstrations does not necessarily include all applicable missions.} 
\label{tab:tech}
\begin{center}       
\begin{tabular}{|c|c|} 
\hline\hline
\rule[-1ex]{0pt}{3.5ex}  \textbf{Technology Gap} & \textbf{Funded Suborbital Demonstrations} \\
\hline\hline\hline
\rule[-1ex]{0pt}{3.5ex}  Large-format, high-quantum-efficiency, solar-blind,  & FIREBall\cite{Kyne20}, INFUSE\cite{Witt23}, \\ photon-counting FUV \& NUV detectors & MOBIUS, SISTINE\cite{Behr23}\\
\hline
\rule[-1ex]{0pt}{3.5ex}  High-throughput, large-format multi-object and & FORTIS\cite{Kutyrev23}, INFUSE\cite{Witt23}, \\ integral field spectroscopy & SUMO\cite{Halferty23} \\
\hline
\rule[-1ex]{0pt}{3.5ex}  High-throughput bandpass selection filters & FIREBall, FLUID\cite{Nell23} \\
\hline
\rule[-1ex]{0pt}{3.5ex}  High-reflectivity FUV-to-NIR coatings & INFUSE\cite{Witt23}, SISTINE\cite{Behr23} \\
\hline
\rule[-1ex]{0pt}{3.5ex}  High-peformance UV gratings & FORTIS\cite{Carlson21}, MOBIUS\\
\hline
\hline
\end{tabular}
\end{center}
\end{table}

Intrinsic to the technology maturation enabled via suborbital development is the ability to encounter and overcome integration and implementation challenges prior to use on higher-cost missions. In the case of the Faint Intergalactic-medium Redshifted Emission Balloon (FIREBall), for example, an emerging technology in high-efficiency, photon-counting UV detector capabilities has been developed and matured at the component level and as part of a full-scale instrument via a balloon payload. 
FIREBall is an Explorer-sized payload that uses a 1-m primary mirror to feed a UV spectrograph that is sensitive over a narrow ($\approx$20~nm) bandpass centered around $\approx$205~nm, where atmospheric transmission is $\approx$50\% at an altitude of $\approx$120~kft\cite{Mat11}.  
FIREBall was originally conceived as a UV fiber-based IFS with a MCP to detect and characterize faint UV emission from the circumgalactic and intergalactic medium (CGM and IGM, respectively)\cite{Grange05}. 

After flying the first UV IFS, a second iteration of FIREBall, FIREBall-2 (FB2), was developed as a multi-object spectrograph (MOS) to increase instrument sensitivity and implement new technology\cite{Grange14}. 
A significant FB2 enabling technology is the instrument's focal-plane camera, built around a UV delta-doped EMCCD developed through a collaboration between Caltech, the Jet Propulsion Laboratory (JPL), and Teledyne e2v. The FB2 EMCCD, which is controlled with a N\"uv\"u CCD Controller for Counting Photons\cite{Daigle18}, provided an increase of $>$5$\times$ higher quantum efficiency (QE) in the FB2 band relative to FB1 and helps contribute to an overall improvement in instrument efficiency of more than an order of magnitude from FB1 to FB2. Further, the detector and controller development through FB2 helped infuse both technologies for the coronagraph instrument on the Roman Space Telescope \cite{Kyne20, Mennesson22}.

Throughout the FB2 program, which now includes two suborbital balloon flights, characteristics related to extended cosmic-ray trails, persistence issues, charge-transfer inefficiency, and a dark current plateau at low temperatures have been identified via integration with the FB2 instrument \cite{Picouet24}. Further, implementation challenges surrounding instrumental background, the need for space-qualified cryocooling, systems-level heat dissipation, power delivery, and telemetry communications all arose as a result of integrating the detector technology in a flight instrument. As a result, significant effort has been invested in addressing various technology-specific challenges to optimize performance and maximize EMCCD sensitivity for faint astronomical sources. The iterative nature of FIREBall and other suborbital projects allows the technology to be optimized for a specific science and instrument application, flown, recovered, and further developed to address both application-specific needs and broadly applicable performance improvements.

\subsection{Personnel development through suborbital programs}
In addition to their science capabilities and use as key development pathways for instruments and technologies, suborbital projects have also established themselves as critical components in training leaders of larger missions. 
An analysis of the NASA astrophysics Explorer-class missions shows that nearly every Explorer principal investigator (PI) previously held an extensive role on at least one suborbital mission (Table \ref{tab:explorers}). 
Further, nearly all of the Explorer science instruments were developed with suborbital heritage to varying degrees, ranging from component-level heritage (e.g. coatings and detector technologies developed for UVEX) to full-scale instrument development (e.g. the COSI instrument\cite{Beechert22}). 

\begin{table}[t]
\caption{A list of the PI-led NASA astrophysics Explorer missions. For each mission, the principal investigator is listed along with whether the instrument and PI had heritage via NASA suborbital programs. }
\label{tab:explorers}
\begin{center}       
\begin{tabular}{|c|c|c|c|c|} 
\hline\hline
\rule[-1ex]{0pt}{3.5ex}  \textbf{Launch Year} & \textbf{Mission} & \textbf{PI} & \textbf{Instrument Heritage} & \textbf{PI Heritage} \\
\hline\hline\hline
\rule[-1ex]{0pt}{3.5ex}  2030 (exp.) & UVEX\cite{Kulkarni21} & Harrison & Rockets \& Balloons & Balloons\cite{Harrison00} \\
\hline
\rule[-1ex]{0pt}{3.5ex}  2027 (exp.) & COSI\cite{Tomsick19} & Tomsick & Balloons\cite{Beechert22} & Balloons\cite{Beechert22} \\
\hline
\rule[-1ex]{0pt}{3.5ex}  2025 & SPHEREx\cite{Korngut18} & Bock & Rockets\cite{Nguyen18} & Rockets\cite{Bock13} \& Balloons\cite{Gualtiere18} \\
\hline
\rule[-1ex]{0pt}{3.5ex}  2021 & IXPE\cite{Weisskopf22} & Weisskopf & Rockets & Rockets\cite{Weisskopf74} \& Balloons\cite{Ramsey93} \\
\hline
\rule[-1ex]{0pt}{3.5ex}  2018 & TESS\cite{Ricker14} & Ricker & No suborbital\cite{ricker} & Balloons\cite{Ricker73}  \\
\hline
\rule[-1ex]{0pt}{3.5ex}  2012 & NuSTAR\cite{Harrison13} & Harrison & Balloons\cite{Harrison05} & Balloons\cite{Harrison00}\\
\hline
\rule[-1ex]{0pt}{3.5ex} 2009 & WISE\cite{Wright09} & Wright & No suborbital & Rockets \& Balloons\cite{Wright77} \\
\hline
\rule[-1ex]{0pt}{3.5ex}  2004 & Swift\cite{Gehrels04} & Gehrels & Rockets\cite{Mendenhall96} \& Balloons\cite{Stahle96} & Balloons\cite{Gehrels83} \\
\hline
\rule[-1ex]{0pt}{3.5ex} 2003 & GALEX\cite{Martin05} & Martin & Rockets\cite{Schiminovich01} \& Balloons\cite{Martin97} & Rockets\cite{Martin89} \\ 
\hline
\rule[-1ex]{0pt}{3.5ex} 2001 & WMAP\cite{Bennett03} & Bennett & Balloons\cite{Staggs96} & No suborbital heritage \\
\hline
\rule[-1ex]{0pt}{3.5ex} 1999 & FUSE\cite{Moos98} & Moos & Rockets\cite{Wilkinson93} & Rockets\cite{Weinstein77} \\
\hline
\rule[-1ex]{0pt}{3.5ex} 1998 & SWAS\cite{Melnick98} & Melnick & No suborbital\cite{melnick} & Airborne\cite{Melnick78} \& Balloons\cite{melnick} \\ 
\hline
\end{tabular}
\end{center}
\end{table}

The outcome of suborbital leaders becoming the PIs of larger missions naturally follows from the nature of NASA's suborbital programs, which emphasize involvement and leadership opportunities for students and other early-career scientists and engineers. Future NASA PIs are just one of the ways in which suborbital projects contribute to education and personnel development, however. FB2, for example, has seen three postdoctoral scholars serve as the day-to-day project and technical leads over the course of two balloon flights, each of whom has now progressed to faculty positions at R1 universities and earned NASA Roman Technology Fellowships\cite{roman}. In addition, several other postdoctoral researchers have contributed to the NASA-funded team as subsystem or integration leads, and at least seven graduate students have used various aspects of FB2 (e.g. technology development, subsystem design and implementation, science development, etc.) as a significant contribution toward earned or soon-to-be earned PhDs and for standalone publications. Placement for the FB2 students and postdocs includes tenure-track faculty appointments, full-time positions at NASA centers, scientist and engineering positions in industry, and non-tenured academic research positions.  The growth of the early-career contributors to FB2 has expanded the collaboration to include two additional domestic universities and three new institutional leads, broadening both the workforce development of the project itself and for contributions to general NASA astrophysics. Similar inclusion and placement of students and early-career researchers is emphasized by many other suborbital projects, and is a key pillar of NASA's suborbital programs in general. 

\subsection{Suborbital projects within the GOMaP}
\label{sec:subs_gomap}

As outlined in Section \ref{sec:intro}, Astro2020 recommended a maturation program to effectively prepare and increase the cadence of Flagship-class NASA missions \cite{Astro2020}. The Panel on Electromagnetic Observations from Space in Astro2020 further described a notional ``grand technology roadmap'' to guide technology development and increase technology readiness levels (TRLs) for key technologies over a short ($\approx$5 years) period of time. This suggested roadmap is aligned with the implemented GOMaP, which seeks to mature component technologies by the end of the current decade \cite{Feinberg24}. Such accelerated technology maturation is consistent with the strengths offered via the suborbital program: rapid, low-cost access to space and the ability to iterate on and space-prove low-TRL technologies. To continue to allow development through suborbital programs and direct that development toward specific mission concepts, the GOMaP could include targeted applications of suborbital rockets and balloons that can further advance observational studies of astrophysical sources, specific technologies identified for use in the Flagship mission, and scientists and engineers working toward contributions for large missions. 

A successful GOMaP could leverage the existing NASA suborbital infrastructure and community to directly advance key initiatives with a relatively small fraction of the budget recommended in Astro2020. 
Further, targeted GOMaP funding for such programs will ensure that the traditional sources of support for astrophysics technology and suborbital instruments continue to enable a wavelength-balanced suite of projects across the full electromagnetic spectrum; since the GOMaP is intended to be a perpetual program that applies to future strategic missions beyond the HWO, traditional funding sources (e.g. the APRA program) will remain critical to the advancement of future missions until they enter the GOMaP pipeline, just as they have for the HWO (refer to Table \ref{tab:tech}). As the astrophysics community looks toward a long-term GOMaP and development of the HWO and future Flagships, suborbital projects can continue to provide a key development pathway toward realizing the future of NASA astrophysics.

\section{Other suborbital and small NASA missions}
\label{sec:smallsats}

Coincident with the rise of sounding rockets and stratospheric balloons, NASA has employed a third suborbital platform since the 1960s: airborne observatories. Such observatories consist of telescopes and instruments that obtain observations of astronomical sources from airplanes. The most recent airborne observatory, SOFIA, flew a 2.5-m telescope at $\approx$14~km to enable observations above water vapor in the Earth's atmosphere \cite{Gehrz09}. SOFIA, like its airborne predecessors, was an impactful observatory for infrared astronomy and flew several different instruments throughout its service life. 
Though SOFIA had a similar flight duration as suborbital balloons (single-night flights lasting 10 hours), airborne missions compose a separate mission class in terms of funding, implementation, and impact on NASA astrophysics development and are not directly comparable to traditional suborbital programs.  

An additional mission type that is often included within the envelope of NASA's suborbital programs is the smallest class of astrophysics missions, CubeSats. Though technically orbital satellites, CubeSats emerged as astrophysics projects via the NASA APRA program and have typically been treated similarly to rockets and balloons in terms of cost, development cycles, and project management. As orbiting payloads, CubeSats offer different advantages to suborbital projects and occupy a unique position along the axes of payload size, complexity, and access to space. Though they are not capable of demonstrating full-system architectures or large instrument components and lack the ability to recover and continue development of a flown technology, CubeSats can offer superior platforms for the in-space demonstration of detectors, coatings, and other technologies that can be implemented fully in a small-format payload. Such projects also offer educational opportunities and experiences for key personnel that go beyond those applicable to many suborbital instruments; different radiation environments, orbital parameters and observing efficiency, long-duration power consumption and thermal considerations, and spacecraft acquisition and integration are just some of the instrument and mission design aspects on which scientists and engineers can gain expertise early in their careers via leadership roles on CubeSats. 

In recent years, however, the relative cost and development time for CubeSats has increased. Most of these missions are now requiring multiple APRA program development cycles to realize launch and operations, with budgets that have increased substantially from the first astrophysics CubeSats. 
Though the first such CubeSat, HaloSat\cite{Kaaret19}, was launched in just three years and the second astrophysics CubeSat, CUTE\cite{France23}, in less than five, subsequent missions have required progressively longer development cycles; each other mission that reached the end of its initial project duration was awarded a continuation program to realize launch and operations, with total development cycles approaching 7-8 years before launch. 
The growing time from conception to launch for CubeSats puts them at a disadvantage relative to suborbital projects in terms of responsiveness to emerging science and technologies, and the ability for students and early-career researchers to experience the full project lifecycle. 

In 2025, several additional CubeSats are expected to join the three NASA-funded astrophysics missions (HaloSat, CUTE, and BurstCube\cite{Perkins20}) that have already launched. Whereas the first CubeSats relied heavily on relatively proven technologies to allow a simpler path to a functional orbital instrument, more recent CubeSat instruments will emphasize emerging technologies to enable their science objectives. SPRITE will seek to raise the TRL of two high-priority UV technologies, enhanced lithium fluoride mirror coatings and a next-generation borosilicate MCP, that enable a low-resolution FUV spectrograph to map ionizing radiation from supernova remnants and nearby galaxies\cite{Fleming22, Indahl23}. SPARCS is another UV CubeSat set to launch in 2025 to monitor M-dwarf activity to study stellar evolution and the impact of stellar activity on exoplanet habitability \cite{Jensen24}. The two-channel SPARCS FUV/NUV instrument will will fly a pair of delta-doped CCDs and, for the first time, a directly deposited metal dielectric filter\cite{Jewell24}. Finally, BlackCAT is an upcoming wide-field X-ray CubeSat instrument to study gamma-ray bursts and other transient events across a broad X-ray bandpass (0.5 - 20~keV) \cite{Falcone24}. In addition to its science capabilities, BlackCAT will be the first orbital mission to incorporate X-ray hybrid CMOS detectors, a technology under development for future large X-ray missions \cite{Colosimo24, Stone24, ABTR24}. If successful, each of these upcoming CubeSats will advance key technologies and achieve new science observations that are not obtainable with currently operating missions. 

Lastly, another relatively new entry in the NASA astrophysics portfolio are SmallSats, which are generally larger than CubeSats but still significantly smaller (and less expensive) than Small Explorer missions. SmallSats are solicited with a \$20M cost cap as part of the NASA Pioneers program. The first astrophysics SmallSats are set to launch in the coming years and include science-focused missions for studying the CGM in the UV\cite{Chung21}, exoplanet atmospheres via transit spectroscopy in the visible-to-IR\cite{Quintana24}, and an all-sky gamma-ray monitor to study neutron star mergers\cite{Woolf24}. Future concepts such as the SmallSat Technology Accelerated Maturation Platform (STAMP), however, seek to incorporate the strengths of the suborbital program (technology and personnel maturation) in SmallSats to directly apply GOMaP development in an orbital platform\cite{France24}. Though such SmallSats are more expensive than suborbital missions and are more limited in instrument size, the STAMP program would allow for on-orbit technology demonstrations and instruments designed for science observations not achievable with currently operating missions. The first STAMP concept mission would advance the space readiness of broadband UV coatings, high-sensitivity UV detectors, and multi-object selection technologies \cite{France24}. Such an approach is consistent with GOMaP objectives and could help reduce the overall cost and development cycle of the flagship missions participating in the program.

As the number of launched astrophysics CubeSats doubles in 2025 and the first SmallSats quickly follow, the reliability and scientific impact of these small satellites on NASA astrophysics will become more clear. Further, the role that CubeSats and SmallSats have within the broader astrophysics portfolio will be better understood: it is not yet clear whether these mission are best applied as low-risk science missions like their larger orbital counterparts, or as more quickly developed maturation tools that can also fulfill niche science objectives (i.e. similar to traditional suborbital). As summarized in Astro2020, it remains to be seen whether small satellites ``will prove to be an effective platform for a range of astrophysics investigations.''\cite{Astro2020}. Despite their relatively recent arrival as core components of NASA's astrophysics portfolio, however, the comparatively inexpensive access to orbital platforms provided by CubeSats and SmallSats offers unique capabilities to astrophysics investigations and development into the future. 
A maturation program that can sufficiently and effectively advance technologies, mission architectures, science, and personnel for the future great observatories will almost certainly include significant contributions from a balanced suite of suborbital projects and small, quickly developed satellites.

\section{Summary}
\label{sec:summary}
Suborbital project have long played a critical role in the NASA mission portfolio for advancing observational studies of various astrophysical phenomena, maturing technologies for larger-class missions, and providing training and leadership opportunities for a broad range of early-career scientists and engineers. Prior to the formulation of the GOMaP, such projects organically became a primary resource for maturing instruments and personnel for inclusion in large NASA missions. Explorer-class missions (traditionally the largest PI-led missions in the NASA astrophysics portfolio) have drawn heavily from personnel and instruments developed on suborbital projects, and strategic astrophysics technologies have likewise been developed largely through suborbital applications. 

Suborbital programs are able to design instruments with more advanced technology than orbital missions owing to their rapid development timelines, increased launch frequency, and higher risk tolerance. These instruments then allow for unique science capabilities that can obtain new observations of various astrophysical sources across a broad range of wavelengths, including wavelengths not accessed with currently operating orbital instruments. As science and supporting technologies continue to evolve and the GOMaP further motivates maturation for large NASA missions, suborbital programs can leverage their ongoing development to directly address specific science, technologies, and instrument architectures in the form of optimized instruments and targeted objectives, all while continuing to train and develop the personnel that will enable future missions.

In more recent years, CubeSats and SmallSats are emerging as contributors to the NASA astrophysics mission portfolio. Though only a few such missions have launched at the time of this review, that number will increase threefold in the following years as more CubeSats achieve orbit and the first collection of SmallSats are launched. As the applicability and utility of these small, orbital programs solidifies in the coming years, there is the potential for them to contribute to the GOMaP and longer-term pathfinding development for the HWO and beyond.

\subsection*{Disclosures}
The author currently and/or has previously benefited from funding directed toward suborbital missions via the NASA Astrophysics Research and Analysis program. 

\subsection* {Acknowledgments}
The author thanks Vincent Picouet for the helpful discussions about the role that suborbital projects play in astrophysics research both domestically and internationally, and Kevin France for helpful comments and insight.  

\bibliography{article}  
\bibliographystyle{spiejour}   

\listoffigures
\listoftables

\end{spacing}
\end{document}